\documentclass[twocolumn,showpacs,preprintnumbers,amsmath,amssymb,pra,footinbib]{revtex4}

\usepackage{graphicx}
\usepackage{bm}
\usepackage{braket}
\usepackage{comment}
\usepackage{amsmath}
\usepackage{amssymb}
\usepackage{amsthm}
\usepackage{subfigure}

\newtheorem{theorem}{Theorem}

\newcommand{\myeqnref}[1]{Eq.~\eqref{#1}}

\newcommand{\mysubsection}[1]{\section{#1}}

\begin{document}

\title{
Getting Information on Independently Prepared Quantum States --- When Are Individual Measurements as Powerful as Joint Measurements?
}

\author{Chi-Hang Fred Fung}
\author{H. F. Chau}
\affiliation{Department of Physics and Center of Theoretical and Computational Physics, University of Hong Kong, Pokfulam Road, Hong Kong}

\begin{abstract}
Given a composite quantum system in which the states of the subsystems are independently (but not necessarily identically) prepared, we construct separate measurements on the subsystems from any given joint measurement such that the former always give at least as large information as the latter.
This construction offers new insights into the understanding of measurements on this type of composite systems.
Moreover, this construction essentially proves the intuition that
separate measurements on the subsystems are sufficient to extract the maximal information about the separately prepared subsystems, thus making a joint measurement unnecessary.
Furthermore, 
our result implies that individual attacks are as powerful as collective
attacks in obtaining information on the raw key in quantum key distribution.
\end{abstract}


\maketitle

\mysubsection{Introduction}
Quantum states can be used to convey information.
A sender, Alice, may prepare a few quantum particles, whose states depend on the message itself, and send them through a quantum channel to the receiver, Bob.
To determine the message, Bob performs a quantum measurement on his received quantum states.
When the quantum states live in multiple quantum subsystems, Bob may perform separate quantum measurements on the subsystems to 
learn about the message.
Alternatively, he may perform a joint quantum measurement on all subsystems together.
In general, performing separate quantum measurements on the subsystems is not powerful enough to extract maximal information on the input state.
In fact, the capacities of certain quantum channels~\cite{Holevo1998,Schumacher1997} and the maximum information that can be extracted from certain unentangled but classically correlated states~\cite{Bennett1999} can only be attained via joint measurements.
But what if each subsystem is independently used to convey information?
Perhaps measuring each subsystem separately is already good enough to extract maximum amount of information on the states in each subsystem.
Here we prove this intuition by explicitly constructing an individual measurement from a given joint measurement such that information gain from former is at least as large as the latter.
This construction offers new insights into the understanding of measurements on this type of composite systems.
Furthermore, we explain the operational meaning of such construction and discuss
its implication to quantum key distribution (QKD).

\medskip
\mysubsection{Precise definition of our problem}
Suppose that there are $K$ subsystems.
For each subsystem~$k$, Alice selects a state indexed by $a_k$ from a set of normalized density matrices $\{ \phi^{(k)}_{a_k}\}$ with probability $p^{(k)}_{a_k}$.
(In other words, $\operatorname{Tr}(\phi^{(k)}_{a_k})=1$ and $\sum_{a_k} p^{(k)}_{a_k}=1$.
Furthermore, we do not limit the the number of subsystems $K$, the Hilbert space dimension of each subsystem and the number of elements in the set $\{ \phi^{(k)}_{a_k} \}$ for each $k$.
These three numbers may well be infinite.)
The state in each subsystem is selected independently but not necessarily identically.
Suppose that Bob uses a particular joint positive operator-valued measure (POVM) to measure the $K$ subsystems.
In general, this POVM may contain elements that are entangled with the $K$ subsystems.
The purpose of this paper is to construct an individual measurement in the $K$ subsystems that can extract no less Shannon mutual information about Alice's states than the original joint POVM.
Here individual measurement refers to the one composed of $K$ independent POVM's each operating on one subsystem.
Thus, by showing that such an individual measurement exists, we confirm the intuition that maximal information on separately prepared subsystems can be extracted separately.

We present two methods for constructing such an individual measurement from the original joint measurement.
Both methods draw on the observation that knowing the states of the other subsystems gives rise to a projected measurement on a subsystem.
The first method is simpler to apply, while the second one admits an intuitive explanation for why it gives at least as large information as the original joint measurement.
We also provide the operational meaning for the second method.

\medskip
\mysubsection{Construction~1}
Without lost of generality, let us consider the case of having two subsystems (that is, $K = 2$) and denote the original joint POVM as $\{M_b: \forall b\}$.
The case of $K > 2$ can be constructed and proven in a similar way.
The key idea of constructing the individual measurement is to focus on measuring one particular subsystem and look at what effective measurement is performed on it.
Thus, let us focus, say, on subsystem~$1$ (${\mathbb S}_1$).
If Alice always prepares subsystem~2 (${\mathbb S}_2$) in the state $\phi$, then the effective measurement on ${\mathbb S}_1$ is
\begin{align}
\label{eqn-effective-measure-1}
\{
\operatorname{Tr}_2 [(\mathbb{I} \otimes \phi) M_b]
: \forall b
\} .
\end{align}
More generally, if Alice may prepare the state in ${\mathbb S}_2$ in more than one way, then for each state sent by Alice in ${\mathbb S}_2$, there corresponds a set of POVM elements similar to \myeqnref{eqn-effective-measure-1}.
The entire POVM is then composed of all these sets.
One may regard the set of POVM elements corresponding to a state in ${\mathbb S}_2$ as the effective measurement on ${\mathbb S}_1$ when Alice sends that state.
Therefore, how likely Bob uses this set of POVM elements should be weighted by the \emph{a priori} probability of the corresponding state being sent.
In summary, the effective POVM for ${\mathbb S}_1$ is $\{ M^{(1)}_{b_1} : \forall b_1\}$ where
\begin{align}
\label{eqn-sep-povm-1}
M^{(1)}_{b_1}=\operatorname{Tr}_2 [(\mathbb{I} \otimes \phi^{(2)}_{a_2}) M_b] p^{(2)}_{a_2} ,
\end{align}
and $b_1 \equiv (a_2,b)$ is the index of the POVM element specifying an input state in ${\mathbb S}_2$ and an element of the original joint POVM.
Using the same argument, the effective POVM $\{ M^{(2)}_{b_2} : \forall b_2\}$ for ${\mathbb S}_2$ is
\begin{align}
\label{eqn-sep-povm-2}
M^{(2)}_{b_2}=\operatorname{Tr}_1 [(\phi^{(1)}_{a_1} \otimes \mathbb{I} ) M_b] p^{(1)}_{a_1}
\end{align}
where $b_2 \equiv (a_1,b)$ is similarly defined.

\begin{theorem}
\label{thm-construction1}
{\rm
Suppose Alice prepares the states of subsystems $k=1,2$ independently.
Then, the amount of Shannon mutual information provided by an individual measurement on the two subsystem using the POVM's $\{ M^{(k)}_{b_k} \}, k=1,2$ whose elements are given in \myeqnref{eqn-sep-povm-1} and \myeqnref{eqn-sep-povm-2} is at least as large as the Shannon mutual information provided by the original joint POVM $\{M_b\}$.
}
\end{theorem}
\begin{proof}
Let the capitalized symbols $A_1$, $A_2$, and $B$ denote the random variables for the input states $a_1$ and $a_2$ and the original joint POVM outcome $b$, respectively.
Since the two subsystems are independent, the mutual information for the original joint POVM $\{M_b\}$ is given by
\begin{align}
I_{1+2} &\triangleq I(A_1, A_2 ; B) = H(A_1, A_2 )-H(A_1, A_2 | B) \nonumber \\
&= H(A_1) + H( A_2 )-H(A_1, A_2 | B) .
\end{align}
Here, the functions $I(\cdot;\cdot)$ and $H(\cdot)$ are the mutual information between its arguments and the entropy of its argument, respectively.
The mutual information between the input and the output of subsystem $k=1,2$ is
\begin{equation}
I_k \triangleq I(A_k; B_k) = H(A_k)-H(A_k | B_k) . \label{eqn-mutual-information_Ik}
\end{equation}
To prove this theorem, it suffices to show that $I_1 + I_2 \geq I_{1+2}$
which can be expressed as
\begin{align}
&H(B_1)-H(A_1 , B_1)+
H(B_2)-H(A_2 , B_2) \nonumber \\
\geq &
H(B)-H(A_1, A_2 , B).
\label{eqn-inequality-to-prove}
\end{align}
We proceed by establishing a crucial relationship between the joint probability of the overall system and that of each subsystem.
The former, with inputs $A_1$ and $A_2$ and output $B$, is given by
\begin{align}
\label{eqn-joint-prob-overall-system}
&\text{Pr} \{A_1=a_1,A_2=a_2,B=b\} \nonumber \\
= &\operatorname{Tr} [(\phi^{(1)}_{a_1} \otimes \phi^{(2)}_{a_2} )M_b]  p^{(1)}_{a_1} p^{(2)}_{a_2},
\end{align}
while the latter, with input $A_k$ and output $B_k$ for subsystem $k$, is given by
\begin{align}
\label{eqn-joint-prob-system1}
\text{Pr} \{A_k=a_k,B_k=b_k\}=\operatorname{Tr} [\phi^{(k)}_{a_k} M^{(k)}_{b_k}] p^{(k)}_{a_k} .
\end{align}
Here, the POVM element $ M^{(k)}_{b_k}$ is given in \myeqnref{eqn-sep-povm-1} or \myeqnref{eqn-sep-povm-2}.
We relate these two probabilities for say ${\mathbb S}_1$ by expanding the POVM element in \myeqnref{eqn-joint-prob-system1} as follows:
\begin{align}
&\operatorname{Pr} \{A_1=a_1,B_1=(a_2,b)\} \nonumber \\
=&\operatorname{Tr}_1 [\phi^{(1)}_{a_1} (\operatorname{Tr}_2 [(\mathbb{I} \otimes \phi^{(2)}_{a_2}) M_b] p^{(2)}_{a_2})] p^{(1)}_{a_1} \nonumber \\
=&\operatorname{Pr} \{A_1=a_1,A_2=a_2,B=b\} .
\label{eqn-joint-prob-rel-1}
\end{align}
This crucial relationship between the probabilities directly translates into 
a relationship 
between the entropies:
\begin{align}
H(A_1 , B_1)&=\sum_{a_1,b_1} f( 
\text{Pr} \{A_1=a_1,B_1=b_1\}
) \nonumber \\
&= \sum_{a_1,(a_2,b)} f( 
\text{Pr} \{A_1=a_1,B_1=(a_2,b)\}
) \nonumber \\
&= \sum_{a_1,a_2,b} f( 
\text{Pr} \{A_1=a_1,A_2=a_2,B=b\}
) \nonumber \\
&=H(A_1,A_2,B)
\end{align}
where $f(x)=-x \log_2 x$.
Replacing ${\mathbb S}_1$ by ${\mathbb S}_2$, we have
\begin{align}
\label{eqn-entropy-A1B1}
H(A_1 , B_1) = H(A_1,A_2,B) = H(A_2 , B_2).
\end{align}
By the same token, we know that
\begin{align}
H(B_1)
&= \sum_{a_2,b} f( \sum_{a_1} 
\text{Pr} \{A_1=a_1,A_2=a_2,B=b\}
) \nonumber \\
&= H(A_2,B)
\label{eqn-entropy-B1}
\end{align}
and
\begin{align}
\label{eqn-entropy-B2}
H(B_2)= H(A_1,B).
\end{align}
 From Eqs.~\eqref{eqn-entropy-A1B1}-\eqref{eqn-entropy-B2},
Eq.~\eqref{eqn-inequality-to-prove} is reduced to the well-known entropy inequality in (classical) information theory~\cite{Cover2006}
\begin{align}
\label{eqn-mutualinfo-ineq3}
H(A_1|B) \geq H(A_1|A_2,B) .
\end{align}
Therefore, this theorem is proved.
\end{proof}

\medskip
\mysubsection{Construction~2}
Recall that the effective POVM for each subsystem (given in \myeqnref{eqn-sep-povm-1} and \myeqnref{eqn-sep-povm-2}) is a mixture of sub-POVM's each corresponding to a state sent in the other subsystem.
Now the key observation is that Bob can use any of these sub-POVM's on one subsystem irrespective of the actual state sent in the other.
That is to say, Bob can use on ${\mathbb S}_2$ the sub-POVM corresponding to one state in ${\mathbb S}_1$ even though Alice has really sent another state in ${\mathbb S}_1$.
Therefore, among all sub-POVM's for a particular subsystem, we can pick the one that provides the highest mutual information.
This sub-POVM, alone, then constitutes the effective POVM for that subsystem.
And this construction results in the effective POVM $\{M^{(1)}_{b}:\forall b\}$ for ${\mathbb S}_1$ where
\begin{align}
\label{eqn-sep-povm-construction2-1}
M^{(1)}_{b}=\operatorname{Tr}_2 [(\mathbb{I} \otimes \phi^{(2)}_{a_2}) M_b] ,
\end{align}
and $\phi^{(2)}_{a_2}$ is chosen to be one of the possible states of 
${\mathbb S}_2$ so that $\{M^{(1)}_{b}\}$ maximizes the mutual information for 
${\mathbb S}_1$.
Similarly, the elements of the effective POVM $\{M^{(2)}_{b}:\forall b\}$ for ${\mathbb S}_2$ are
\begin{align}
\label{eqn-sep-povm-construction2-2}
M^{(2)}_{b}=\operatorname{Tr}_1 [(\phi^{(1)}_{a_1} \otimes \mathbb{I}) M_b] ,
\end{align}
where
$\phi^{(1)}_{a_1}$ is chosen to be one of the possible states of 
${\mathbb S}_1$ so that $\{M^{(2)}_{b}\}$ maximizes the mutual information for 
${\mathbb S}_2$.

\begin{theorem}
\label{thm-construction2}
{\rm
Suppose Alice prepares the states of subsystems ${\mathbb S}_1, {\mathbb S}_2$ independently.
Then, the amount of Shannon mutual information provided by an individual measurement 
using the POVM's whose elements are defined in \myeqnref{eqn-sep-povm-construction2-1} for ${\mathbb S}_1$
and
in \myeqnref{eqn-sep-povm-construction2-2} for ${\mathbb S}_2$
is at least as large as the Shannon mutual information provided by the original joint POVM $\{M_b\}$.
}
\end{theorem}
\begin{proof}
We focus on ${\mathbb S}_1$ as the case of ${\mathbb S}_2$ is similar.
It suffices to show that the mutual information for the POVM whose elements are defined in \myeqnref{eqn-sep-povm-construction2-1} is no less than that in \myeqnref{eqn-sep-povm-1}, and invoke Theorem~\ref{thm-construction1}.
Observe that
\begin{align}
I(A_1;B_1)
&= H(A_1)-[H(A_1 , B_1)-H(B_1)] \nonumber \\
&= H(A_1)-[H(A_1 , A_2,B)-H(A_2,B)] \nonumber \\
&= H(A_1|A_2)-[H(A_1 , B|A_2)-H(B|A_2)]  \nonumber \\
&= \sum_{a_2} p^{(2)}_{a_2} I(A_1;B|A_2=a_2) \nonumber \\
&\leq \max_{a_2} I(A_1;B|A_2=a_2) ,
\label{eqn-sep-povm-construction2-weighted1}
\end{align}
where the second line is due to Eqs.~\eqref{eqn-entropy-A1B1} and \eqref{eqn-entropy-B1},
and the third line is due to the fact that the states in the two subsystems are independent and that one can arbitrarily add and subtract $H(A_2)$.
We proceed to verify that $I(A_1;B|A_2=a_2)$ is indeed the mutual information for the sub-POVM consisting of elements given in \myeqnref{eqn-sep-povm-construction2-1}.
The probability of observing outcome $b$ with input $A_1=a_1$ corresponding to \myeqnref{eqn-sep-povm-construction2-1} equals
$\operatorname{Tr}_1 [\phi^{(1)}_{a_1} (\operatorname{Tr}_2 [(\mathbb{I} \otimes \phi^{(2)}_{a_2}) M_b] )] p^{(1)}_{a_1} = \text{Pr} \{A_1=a_1,B=b|A_2=a_2\}$.
This means that the corresponding mutual information for this POVM is $I(A_1;B|A_2=a_2)$.
Therefore, \myeqnref{eqn-sep-povm-construction2-weighted1} shows that indeed the POVM of \myeqnref{eqn-sep-povm-1} can be broken down into sub-POVM's each corresponding to one value of $a_2$.
Thus, when Bob always uses the sub-POVM corresponding to the $a_2$ that maximizes $I(A_1;B|A_2=a_2)$, the resulting mutual information is no less than that of using the weighted average of the sub-POVM's.
\end{proof}

\medskip
\mysubsection{Intuitive explanation of Construction~2
}
\begin{figure}
\subfigure[Introducing the phantom subsystem ${\mathbb P}_2$ in which a joint measurement is performed on it and the real subsystem ${\mathbb S}_1$.]{
\makebox[\columnwidth]{
\includegraphics[scale=.75]{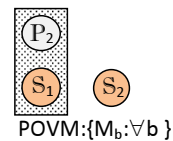}\label{fig-virtual-a}}}
\\
\subfigure[Alice reveals the state in ${\mathbb P}_2$, allowing a projected measurement to be performed on ${\mathbb S}_1$.]{
\makebox[\columnwidth]{
\includegraphics[scale=.75]{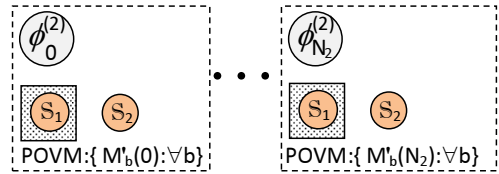}\label{fig-virtual-b}}}
\\
\subfigure[Bob uses the best projected measurement for ${\mathbb S}_1$.]{
\makebox[\columnwidth]{
\includegraphics[scale=.75]{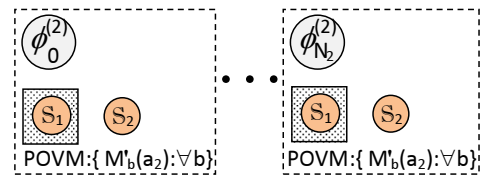}\label{fig-virtual-c}}}
\\
\subfigure[${\mathbb P}_2$ is now superfluous and is discarded.]{
\makebox[\columnwidth]{
\includegraphics[scale=.75]{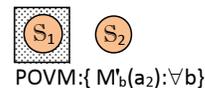}\label{fig-virtual-d}}}
\caption{\label{fig:virtual}
Intuitive explanation for Construction~2, focusing on real subsystem~$1$ (${\mathbb S}_1$).
In each step, the information obtained on ${\mathbb S}_1$ is non-decreasing.
Here, 
dark (light) circles represent real (phantom) subsystems, 
shaded rectangles represent measurements, and $M_b'(x)=
\operatorname{Tr}_2 [(\mathbb{I} \otimes \phi^{(2)}_{x}) M_b]$ represents the POVM element of the projected measurement.
}
\end{figure}
Let us introduce two phantom subsystems (intended to be thrown away later) in addition to the two real subsystems.
The phantom subsystem ${\mathbb P}_k$ serves to replicate real subsystem ${\mathbb S}_k$ for $k = 1,2$ in the sense that they share the same set of states in which Alice may send with the same \emph{a prior} probabilities.
Nevertheless, they are independent of each other and of other subsystems.
Since all the states sent by Alice in the four subsystems are independent, the pair consisting of ${\mathbb S}_1$ and ${\mathbb S}_2$ and the pair consisting of ${\mathbb S}_1$ and ${\mathbb P}_2$ appear to be identical to Bob.
Thus, the amount of information Bob can learn about ${\mathbb S}_1$ from measuring the first pair and that from measuring the second pair using the same joint measurement must be same.
Because of this, we may consider that Bob performs the joint measurement on ${\mathbb S}_1$ and ${\mathbb P}_2$ (see Fig.~\ref{fig-virtual-a}).

Now suppose that Alice tells Bob exactly which state was sent in ${\mathbb P}_2$ (and we will show that delaying this announcement indefinitely turns out to have no bearing on Bob).
Using this extra piece of information, Bob can pick the corresponding POVM elements that are consistent with the phantom state and project it onto ${\mathbb S}_1$ as a measurement operator (see Fig.~\ref{fig-virtual-b}).
Interestingly, this projected measurement turns out to be the effective measurement we have constructed in \myeqnref{eqn-sep-povm-construction2-1} for various values of $a_2$.
Essentially, for each state in ${\mathbb P}_2$ announced by Alice, there corresponds an effective POVM for ${\mathbb S}_1$.
Clearly, with the aid of the extra information in the state of ${\mathbb P}_2$, Bob's information on ${\mathbb S}_1$ in this case is at least as large as that could be obtained with the original joint measurement on the two systems when Alice did not disclose the state of ${\mathbb P}_2$.

Now the key point is that Bob can use any of these effective POVM's on ${\mathbb S}_1$ irrespective of the actual state sent in ${\mathbb P}_2$.
This is because Bob's information on ${\mathbb S}_1$ obtained from using a particular effective POVM does not depend on the state of ${\mathbb P}_2$ as ${\mathbb P}_2$ and ${\mathbb S}_1$ are independent.
Therefore, we can regard that Bob always ignores Alice's announcement of the state in ${\mathbb P}_2$ and uses the effective POVM on ${\mathbb S}_1$ that gives him the maximum amount of information (see Fig.~\ref{fig-virtual-c}).
When Bob always uses only one effective POVM on ${\mathbb S}_1$, the existence of ${\mathbb P}_2$ is irrelevant and thus we can completely discard ${\mathbb P}_2$ along with the announcement of its state (see Fig.~\ref{fig-virtual-d}).
Since Bob always uses the best effective POVM on ${\mathbb S}_1$, the amount of information he gets on ${\mathbb S}_1$ is at least as large as that when he chooses the POVM based on Alice's announcement, which we have already argued is no worse than that when he uses the original joint measurement.

We repeat the previous argument on the pair ${\mathbb S}_2$ and ${\mathbb P}_1$
 to obtain the best effective POVM for ${\mathbb S}_2$.
Finally, the independence of ${\mathbb S}_1$ and ${\mathbb S}_2$ allows us to conclude that
using the best effective POVM for each of them gives 
no less
information on both as the original 
joint POVM.

\medskip
\mysubsection{Multipartite systems}
Our results for the bipartite case given by Theorems~\ref{thm-construction1} and~\ref{thm-construction2} can easily be extended to the multipartite case (including the case of an infinite number of subsystems).
In particular, the POVM elements corresponding to Theorem~\ref{thm-construction1} for a $K$-partite system are
\begin{align}
M^{(k)}_{b_k}&=\operatorname{Tr}_{\ell \neq k} \left[\bigotimes_{\ell=1}^K \phi^{(\ell,k)}_{a_\ell} M_b \right] \prod_{\ell\neq k} p^{(\ell)}_{a_\ell} \: \: \: \forall k ,
\end{align}
where $\phi^{(\ell,k)}_{a_\ell} = \phi^{(\ell)}_{a_\ell}$ if $\ell\neq k$ and $\phi^{(\ell,k)}_{a_\ell} = {\mathbb I}$ otherwise.

\medskip
\mysubsection{Implication to quantum key distribution}
The result in this paper sheds some light on
the various types of eavesdropping attacks in QKD~\cite{Bennett1984,Ekert1991}.
In most QKD protocols such as the famous BB84 protocol \cite{Bennett1984}, 
a legitimate party (Alice) sends a sequence of quantum states each independently chosen from a set of states to another legitimate party (Bob) through a hostile channel controlled by an eavesdropper (Eve).
The goal of Alice and Bob is to derive a secret key from Alice's states and Bob's states.
Eve, on the other hand, attempts to steal their secret by launching an eavesdropping attack.
Two types of keys can be distinguished: the raw key and the final secret key.
Alice's raw key is the bit string corresponding to the quantum states she sends to Bob;
whereas Bob's raw key corresponds to his measurement results on the received qubits
\footnote{We do not consider noisy pre-processing here. See Refs.~\cite{Csiszar1978,Kraus2005,Renner2005} for discussions on noisy pre-processing.}.
Their raw keys may not be secure and error-free; and they derive their final keys from their raw keys via privacy amplification.
For QKD protocols in which Alice sends out independent states (such as BB84~\cite{Bennett1984}, 
SARG04~\cite{Scarani2004}, and
Gaussian-modulated coherent states QKD~\cite{Grosshans2003}),
Eve's probes become independent and 
our result in this paper implies
that individual attacks are as powerful as collective attacks in obtaining information on Alice's raw key\footnote{We assume that Eve only uses her probes and Alice's announcement of the basis information to learn about Alice's raw key; in particular, she does not make use of the error correction information that Alice may reveal publicly.  Note that error correction information may be transmitted by Alice in encrypted form, as is the case of Koashi's security proof~\cite{Koashi2005b}.  In this case, Eve is unable to use the error correction information.}.
In contrast, Smith~\cite{Smith2007} shows that when the key generation rate is concerned, collective attacks are strictly more powerful.
This makes sense since
privacy amplification 
correlates Alice's raw keys in order to obtain the final secret key.

\medskip
\mysubsection{Conclusions}
We show that individual measurement is sufficient to obtain optimal
amount of information on the states in which each subsystem is prepared
independently but not necessary identically based on the observation that
knowing the state of the other subsystems gives rise to a projected
measurement on a subsystem.
Applying our result to the QKD setting shows that individual and collective
attacks are equally powerful in obtaining information on the raw key.
Our work uses Shannon mutual information as the information measure.

We note that Wootters has proved the same result as ours that 
the accessible information is additive for independently prepared subsystems~\cite{DiVincenzo2002b}.
However, implementing his proof idea will result in an ensemble of individual measurements 
(each with a fixed probability of being drawn)
for each subsystem.
In contrast, both our construction methods lead to a single measurement for each subsystem.

Constructing individual measurements from a joint one giving at least as large information in terms of other information measures may be possible.
For example, 
the Csisz\'{a}r measure (see, e.g., \cite{Ferentinos1981,Zografos1986}), which is a generalization of Shannon information, allows such a construction in a special case\footnote
{
The Csisz\'{a}r measure (see, e.g., \cite{Ferentinos1981,Zografos1986})
is defined as
$I^C (f,g) = \sum_x g(x) \Phi\left(\frac{f(x)}{g(x)}\right)$
where $\Phi$ is convex.
Shannon mutual information $I(X;Y)$ is recovered with $I(X;Y)=I^C(p(x)p(y),p(x,y))$ and $\Phi(x)=-\log x$.
Keeping the Csisz\'{a}r measure in the form $I^C(p(x)p(y),p(x,y))=:I^C(X;Y)$, 
if $\Phi$ satisfies the condition
$\Phi(x)+\Phi(y) \geq \Phi(xy)$ 
with $x,y\geq0$,
then we have
$I^C(A_1;E|A_2) + I^C(A_2;E) \geq I^C(A_1, A_2; E)$.
By borrowing
techniques in our paper and also that due to Wootters,
it is not difficult to construct the desired individual measurements for the Csisz\'{a}r measure
from this inequality.
}.

\begin{acknowledgments}
We thank Debbie W. Leung, Hoi-Kwong Lo, Xiongfeng Ma, Marco Piani, and Ming-Yong Ye for enlightening discussions. 
This work is supported by the RGC
grant No. HKU~701007P of the HKSAR Government and 
the Postdoctoral Fellowship program of NSERC of Canada.
\end{acknowledgments}

\bibliographystyle{apsrev}
\bibliography{paperdb}

\end{document}